\documentstyle[12pt]{article}

\def\be{\begin{equation}}
\def\ee{\end{equation}}
\def\bea{\begin{eqnarray}}
\def\eea{\end{eqnarray}}
\def\real{\hbox{{I}\kern-.2em{\bf R}}}

\begin{document}

\title{ROTATING 5D-KALUZA-KLEIN SPACE-TIMES FROM INVARIANT TRANSFORMATIONS}
\author{Tonatiuh Matos$^a$, Dar{\'\i}o N{\'u}{\~n}ez$^b$, Gabino Estevez$^a$,
Maribel Rios$^c$ \\
$^a$Departamento de F{\'\i}sica, \\
Centro de Investigaci{\'o}n y Estudios Avanzados del I. P. N.,\\
A. P. 14-700, 07000 M{\'e}xico, D.F.,MEXICO\\
$^b$Instituto de Ciencias Nucleares, \\
Universidad Nacional Aut{\'o}noma de M{\'e}xico\\
A. P. 70-543, 04510 M{\'e}xico, D. F., MEXICO \\
$^c$ Instituto de F{\'\i}sica y Matem{\'a}ticas \\
Universidad Michoacana de San Nicol{\'a}s de Hidalgo \\
Apdo. Postal 2-82, 58040 Morelia, Michoac{\'a}n, M{\'e}xico}
\date{\today}
\maketitle

\begin{abstract}
Using invariant transformations of the five-dimensional Kaluza-Klein (KK)
field equations, we find a series of formulae to derive axial symmetric
stationary exact solutions of the KK theory starting from static ones. The
procedure presented in this work allows to derive new exact solutions up to
very simple integrations. Among other results, we find exact rotating
solutions containing magnetic monopoles, dipoles, quadripoles, etc., coupled
to scalar and to gravitational multipole fields.
\end{abstract}




\section{Introduction}

In recent years, dilaton fields have been proposed as a strong candidate for
describing dark matter. At a cosmological level it has been used to explain
the Large Scale Structure of the Universe \cite{urena}\cite{DMH}. At a
galactic level, scalar fields seem to play a crucial role in explaining the
curves of rotational velocities vs. radius, observed in all the galaxies 
\cite{fco}\cite{GMV}, and it is thought that it will also play an important
role in several physical phenomena at a local level \cite{dam2}\cite{dam}, $%
i.e.,$ in the realm of compact objects.

From a theoretical point of view, dilaton fields coupled to Einstein-Maxwell
fields, naturally appear in the low energy limit of string theory, and as a
result of a dimensional reduction of the Kaluza-Klein Lagrangian. Therefore,
the study of the Einstein-Maxwell-Dilaton theory is of importance to
investigate the properties of compact objects involving these fields and for
the understanding of more general theories.

On the other hand, if the scalar fields are so important in physics, why
they have not been yet detected? As we just mentioned, in several models
they play an important role, but due to the fact that they interact very
weakly with matter \cite{dam,brena}, in most of the observational tests the
results at most just can not exclude them. It is expected that the scalar
fields will have an important measurable signature of their presence in
regions with strong gravitational fields \cite{esp}. Thus, it is necessary
to have exact analytical solutions to the Einstein-Maxwell-Dilaton theory,
not only perturbative solutions, and then compare with the observations the
predictions made using those exact solutions. The problem with this
approach, is that the field equations are very complicated to be solved
exactly and one must recur to mathematical methods which usually prove to be
very cumbersome. In this work we want to give some simple formulas which
allow us to derive exact rotating dilatons solutions, starting from static
ones, and which avoids many of the mathematical difficulties usually
encountered in deriving exact solutions from seed ones. In this work we
derive three expressions which can be used to genereate families of
solutions, starting from known seed ones. Out of these expressions, ony one
has been previously obtained in reference \cite{TC}

In order to do so, let us start from the Lagrangian 
\begin{equation}
{\cal L}=\sqrt{-g}[-R+2(\nabla \phi )^{2}+e^{-2\alpha \phi }F^{2}]  \label{L}
\end{equation}
This Lagrangian contains very interesting limits. For $\alpha ^{2}=3$
Lagrangian (\ref{L}) contains the Kaluza-Klein theory; for $\alpha ^{2}=1$,
equation (\ref{L}) represents the effective Lagrangian for the low energy
limit of super-strings theory; finally, equation (\ref{L}) contains the
Einstein-Maxwell theory with a minimally couple scalar field for $\alpha
^{2}=0$. This Lagrangian is also very convenient because after a conformal
transformation of the metric, one can obtain a equivalent Lagrangian for an
almost arbitrary scalar-tensor theory of gravity \cite{dam}(with a non
trivial electromagnetic-scalar interaction which can be avoided setting $%
F^{2}=0$). The field equations derived from Lagrangian (\ref{L}) are give by 
\begin{eqnarray}
\nabla _{\mu }(e^{-2\alpha \phi })F^{\mu \nu } &=&0;  \nonumber \\
\nabla ^{2}\phi +{\frac{\alpha }{{2}}}e^{-2\alpha \phi }F^{2} &=&0; 
\nonumber \\
R_{\mu \nu } &=&2\nabla _{\mu }\nabla _{\nu }\phi +2e^{-2\alpha \phi
}(F_{\mu \rho }{F_{\nu }}^{\rho }-{\frac{1}{{2}}}g_{\mu \nu }e^{-2\alpha
\phi }F^{2}).  \label{eqL}
\end{eqnarray}

There exist several exact solutions of equations (\ref{eqL}) (see reference 
\cite{ma2} for solutions with $\alpha ^{2}=3$ and \cite{MNQ} for their
generalization to $\alpha $ arbitrary). Some of them could be models for the
exterior space-time of an astrophysical compact object \cite{ma,brena} or of
a black hole with a scalar field interaction \cite{HH1,HH2,MNQ}. Let us give
two examples of such metrics. The first space-time we deal with behaves
gravitationally like the Schwarzschild solution for $\alpha \neq 0$ and
contains an arbitrary magnetic field. This metric reads \cite{ma2} 
\[
ds^{2}=e^{2k_{s}}\hbox{g}^{\gamma }{\frac{dr^{2}}{1-{\frac{2m}{r}}}}+%
\hbox{g}^{\gamma }\ r^{2}(e^{2k_{s}}d\theta ^{2}+\sin ^{2}\theta \ d\varphi
^{2})-{\frac{1-{\frac{2m}{r}}}{\hbox{g}^{\gamma }}}\ dt^{2} 
\]
\begin{equation}
A_{03,z}=Q\rho \tau _{,z}\ \ ,\ \ \ \ A_{03,\overline{z}}=-Q\rho \tau _{,%
\overline{z}}\ \ ,\ \ \ \ e^{-2\alpha \phi _{0}}={\frac{k_{1}^{2}}{(1-{\frac{%
2m}{r}})\hbox{g}^{\beta }}}  \label{sol}
\end{equation}
where a subindex $0$ stands for a seed solution and 
\[
\hbox{g}=a_{1}\tau +1,\ \ \hbox{and}\ \ e^{2k_{s}}=\left( 1+\frac{m^{2}\sin
^{2}\theta }{r^{2}(1-\frac{2m}{r})}\right) ^{-1/\alpha ^{2}}, 
\]
\noindent In this work we use the coordinates $z=\rho +i\ \zeta =\sqrt{%
r^{2}-2mr}\ \sin \theta +i\ (r-m)\cos \theta $. {\bf A}$_{0}=A_{0\mu
}dx^{\mu },\ $(with $\mu =1...4),$ is the electromagnetic four potential, $m$
the mass parameter, $\gamma =2/(1+\alpha ^{2})$, $\beta =2\alpha
^{2}/(1+\alpha ^{2})$; $Q$ and $a_{1}$ are constants related by 
\[
2\gamma a_{1}^{2}-k_{1}^{2}Q^{2}=0 
\]
Solution (\ref{sol}) can be interpreted as a magnetized Schwarzschild
solution in dilaton gravity for $\alpha \neq 0$. For $\alpha =0$ the
construction of dipoles is different and the form of the metric is not
similar to the Schwarzschild solution any more \cite{ma}. In what follows,
we will assume $\alpha \neq 0$. The function $\tau =\tau (\rho ,\zeta )$ is
a harmonic parameter in a two dimensional flat space, $i.e.$, it is a
solution of the Laplace equation

\begin{equation}
{\frac{1}{2{\rho }}}[(\rho \tau _{,z})_{,\bar{z}}+(\rho \tau _{,\bar{z}%
})_{,z}]=\tau _{\rho \rho }+{\frac{1}{{\rho }}}\tau _{,\rho }+\tau _{\zeta
\zeta }=0.  \label{lap}
\end{equation}
This metric represents the exterior field of a gravitational object with an
arbitrary magnetic field coupled to a scalar field. The metric is singular
for $r=2m$ and for an interior radius determined by the magnetic field. For
a pulsar with magnetic and scalar fields in the region where $r>2m$, this
metric could be an static model of an astrophysical object with magnetic and
scalar fields, metric (\ref{sol}) is always regular in that region.

\bigskip The second metric we will deal with is given by \cite{MNQ} 
\begin{equation}
ds^{2}=\frac{1}{f_{0}}\left[ e^{2k_{0}}\left( d\rho ^{2}+d\zeta ^{2}\right)
+\rho ^{2}d\varphi ^{2}\right] -f_{0}dt^{2},
\end{equation}
where 
\begin{eqnarray}
f_{0} &=&{\frac{e^{\lambda }}{({a_{1}\Sigma _{1}+a_{2}\Sigma _{2})}^{\gamma }%
}},  \nonumber \\
\ e^{-2\alpha \phi _{0}} &=&\kappa _{0}^{2}=\kappa _{1}^{2}({a_{1}\Sigma
_{1}+a_{2}\Sigma _{2}})^{\beta }e^{\lambda -\tau _{0}\tau },  \nonumber \\
\,\ 2A_{04} &=&\psi _{0}={\frac{a_{3}\Sigma _{1}+a_{4}\Sigma _{2}}{{%
a_{1}\Sigma _{1}+a_{2}\Sigma _{2}}}},  \label{manq}
\end{eqnarray}
where $a_{1},...,\kappa _{1}$, and $\tau _{0}$ are constants and $\beta
,\gamma $ are again functions of $\alpha $ defined as $\gamma =2/(1+\alpha
^{2})$, $\beta =2\alpha ^{2}/(1+\alpha ^{2})$; $\tau =\tau (\rho ,\zeta )$
and $\lambda =\lambda (\rho ,\zeta )$ are harmonic functions which satisfy
again the Laplace equation (observe that we have defined a new parameter $%
\lambda $ with respect to the one defined in \cite{MNQ}). $\Sigma _{1}$ and $%
\Sigma _{2}$ are functions given in terms of $\tau $ and the equation (\ref
{manq}) contains two subclasses determined by the functions $\Sigma _{1}$
and $\Sigma _{2}$. For the first subclass we have $\tau _{0}=0$ and 
\begin{equation}
\Sigma _{1}=\tau \;\;\;\;\Sigma _{2}=1  \label{manq1}
\end{equation}
with the relation between the constants 
\begin{equation}
4{a_{1}}^{2}-\kappa _{1}^{2}(1+\alpha ^{2})(a_{1}a_{4}-a_{2}a_{3})^{2}=0.
\label{rest1}
\end{equation}

For the second subclass we have 
\begin{equation}
\Sigma _{1}=e^{q_{1}\tau } \;\;\;\;\Sigma _{2}=e^{q_{2}\tau },  \label{manq2}
\end{equation}
where $q_{1}, q_{2}$ are constants and $\tau _{0}$ satisfy the relation $%
\tau _{0}=q_{1}+q_{2}$; the condition for the constants in this case is
given by: 
\begin{equation}
4 a_{1}a_{2}+{\kappa _{1}}^{2}(1+\alpha ^{2})(a_{1}a_{4}-a_{2}a_{3})^{2}=0.
\label{manq2a}
\end{equation}

One of he most interesting solution contained in this class is the
Gibbons-Maeda \cite{gibb} black-hole 
\begin{equation}
ds^{2}=\hbox{g}^{\gamma }{\frac{dr^{2}}{1-{\frac{2m}{r}}}}+\hbox{g}^{\gamma
}\ r^{2}(e^{2k_{s}}d\theta ^{2}+\sin ^{2}\theta \ d\varphi ^{2})-{\frac{1-{%
\frac{2m}{r}}}{\hbox{g}^{\gamma }}}\ dt^{2}  \label{ghs}
\end{equation}
where 
\begin{eqnarray*}
e^{2\alpha \phi _{0}} &=&\left( 1+{\frac{r_{-}}{{r}}}\right) ^{\frac{%
-2\alpha ^{2}}{{1+\alpha ^{2}}}} \\
\hbox{g} &=&\left( 1+{\frac{r_{-}}{{r}}}\right) ; \\
r_{-}-r_{+} &=&-2m
\end{eqnarray*}
(we have used $r\rightarrow r+r_{-}$ from the original solution). This
metric could represent a static charged black hole containing a scalar field 
$\phi _{0}$ (for a study of this metric see \cite{HH1,HH2}). Observe that in
both metrics (\ref{sol}) and (\ref{ghs}), the space-time is qualitatively
different only for $\alpha =0$, but for $\alpha \neq 0$ the qualitative
behavior is very similar for any $\alpha $ and many of the main features of
the metrics can be obtained for a specific $\alpha $.

On the other hand, real astrophysical objects rotate. For an object like a
pulsar, taking into account the rotation is very important in order to
understand its space-time configuration. Therefore, if we want to model an
astrophysical object with scalar field, we must find the corresponding
rotating metrics of (\ref{sol}) and (\ref{ghs}) to obtain the solution which
we want to use for modeling them.

Using the potential space formalism for 5D gravity introduced by Neugebauer 
\cite{DoctorN}, we were able to find a set of formulas valid for $\alpha
^{2}=3$ in order to obtain rotating exact solutions from a static one, and
without making any complicated integration. We will introduce this formalism
in section two.

In section three we start from the axial-symmetric stationary field
equations derived from equation (\ref{L}) for the specific case $\alpha
^{2}=3$. Using the formalism mentioned above, we will derive new solutions
in section four, and finally show that the Kerr space-time, the
Gibbons-Maeda \cite{gibb}, the Frolov-Zelnikov \cite{FZ} and also their NUT
generalizations are special cases of one of these new solutions.

\section{The Potential-Space Formalism}

For $\alpha ^{2}=3$ the field equations can be derived from a
five-dimensional space-time action. We will deal with a five-dimensional
space-time possessing a Killing vector field $X$ with close orbits. We will
work with stationary space-times, this symmetry implies the existence of a
second Killing vector field $Y$ with close orbits as well. Thus, we start
with a five-dimensional space-times possessing two commuting Killing vectors
fields; a space-like one $X$, representing the inner symmetry and a
time-like one $Y$, representing stationarity. The potential formalism
consists in defining covariantly five potentials in terms of the Killing
vectors $X$ and $Y$. The five potentials are given by \cite{DoctorN} 
\begin{eqnarray}
I^{2} &=&\kappa ^{4/3}=X_{A}X^{A};\,\,\ \
\,\,\,f=-IY_{A}Y^{A}+I^{-1}(X^{A}Y_{A})^{2}  \nonumber \\
\psi  &=&-I^{-2}X_{A}Y^{A};\,\ \ \ \ \ \ \epsilon _{,A}=\epsilon
_{ABCDE}X^{B}Y^{B}X^{D;E}  \nonumber \\
\chi _{,\alpha } &=&-\epsilon _{ABCDE}X^{B}Y^{B}X^{D;E}  \label{expo}
\end{eqnarray}
($A,B,..=1...5$), where $f,\epsilon ,\psi ,\chi $ and $\kappa $ respectively
are the gravitational, rotational, electrostatic, magnetostatic and scalar
potentials; $\epsilon _{ABCDE}$ is the five-dimensional Levi-Civita
pseudo-tensor, $X=X^{A}\partial /\partial x^{A}=\partial /\partial x^{5},$
and $Y=Y^{A}\partial /\partial x^{A}=\partial /\partial t$. We will work
with spaces possessing axial symmetry as well, which is a realistic
assumption for a star. Thus for the axial symmetric stationary case we have
another Killing vector $Z=Z^{A}\partial /\partial x^{A}=\partial /\partial
\varphi $, representing this symmetry. The field equations (\ref{eqL}) in
terms of the five potentials $\Psi ^{A}=(f,\epsilon ,\psi ,\chi ,\kappa )$
read \cite{TM1,MNQ} 
\begin{eqnarray}
\hat{D}^{2}\,\kappa +\left( {\frac{\hat{D}{\,\rho }}{\rho }}-{\frac{\hat{D}{%
\,\kappa }}{\kappa }}\right) \,\hat{D}\kappa +{\frac{{3\,\kappa ^{3}}}{{4\,f}%
}}(\hat{D}\psi ^{2}-{\frac{1}{{\kappa ^{4}}}}\,\hat{D}\,\chi ^{2}) &=&0, 
\nonumber \\
\hat{D}^{2}\,\psi +\left( {\frac{{\hat{D}\,\rho }}{\rho }}+{\frac{{2\,\hat{D}%
\,\kappa }}{\kappa }}-{\frac{{\hat{D}\,f}}{{f}}}\right) \,\hat{D}\,\psi -{%
\frac{1}{{\kappa ^{2}\,f}}}\,(\hat{D}\,\epsilon -\psi \,\hat{D}\,\chi )\,%
\hat{D}\,\chi  &=&0,  \nonumber \\
\hat{D}^{2}\,\chi +\left( {\frac{{\hat{D}\,\rho }}{\rho }}-{\frac{{2\,\hat{D}%
\,\kappa }}{\kappa }}-{\frac{{\hat{D}\,f}}{{f}}}\right) \,\hat{D}\,\chi +{%
\frac{{\kappa ^{2}}}{{f}}}\,(\hat{D}\,\epsilon -\psi \,\hat{D}\,\chi )\,\hat{%
D}\,\psi  &=&0,  \nonumber \\
\hat{D}^{2}\,f+\left( {\frac{{\hat{D}\,\rho }}{\rho }}-{\frac{{\hat{D}\,f}}{{%
f}}}\right) \,\hat{D}\,f+{\frac{1}{{f}}}(\hat{D}\,\epsilon -\psi \,\hat{D}%
\,\chi )^{2}-{\frac{{\kappa ^{2}}}{{2}}}\,\left( \hat{D}\psi ^{2}+{\frac{1}{{%
\kappa ^{4}}}}\,\hat{D}\,\chi ^{2}\right)  &=&0,  \nonumber \\
\hat{D}^{2}\,\epsilon -\hat{D}\,\psi \,\hat{D}\,\chi -\psi \,\hat{D}^{2}\chi
+\left( {\frac{{\hat{D}\,\rho }}{\rho }}-{\frac{{2\,\hat{D}\,f}}{{f}}}%
\right) \,(\hat{D}\,\epsilon -\psi \,\hat{D}\,\chi ) &=&0\,\,\,
\label{fielde}
\end{eqnarray}
where $\hat{D}$ is the differential operator $\hat{D}=(\partial _{\rho
},\partial _{\zeta })$. The field equations (\ref{fielde}) can be derived
from the Lagrangian \cite{DoctorN,ma,MNQ} 
\begin{equation}
{\cal L}={\frac{\rho }{{2\,f^{2}}}}[f_{,i}f^{,i}+(\epsilon _{,i}-\psi \chi
_{,i})(\epsilon ^{,i}-\psi \chi ^{,i})]+{\frac{\rho }{{2\,f}}}\left( \kappa
^{2}\psi _{,i}\psi ^{,i}+{\frac{1}{{\kappa ^{2}}}}\chi _{,i}\chi
^{,i}\right) -{\frac{2\rho }{{3\,\kappa ^{2}}}}\kappa _{,i}\kappa ^{,i},
\label{lag1}
\end{equation}
with $i=(\rho ,\zeta )$. The next step is to look for the invariant
transformations of Lagrangian (\ref{lag1}), which were found in \cite{TM3}.
The invariance group of the Lagrangian (\ref{lag1}) is $SL(3,%
\hbox{{I}\kern-.2em{\bf R}})$. We can write these transformations in a very
simple form as 
\begin{equation}
h\,\,\rightarrow \,\,Ch_{0}C^{T},  \label{tinv}
\end{equation}
where $h$ and $C$ are elements of $SL(3,\hbox{{I}\kern-.2em{\bf R}})$. One
parameterization of matrix $h$ is given by \cite{TM1,ma} 
\begin{equation}
h=-{\frac{1}{{f\,\kappa ^{2/3}}}}\pmatrix{f^2+\epsilon^2-f\,\kappa^2\,%
\psi^2&-\epsilon&-\epsilon\chi+f\,\kappa^2\,\psi\cr -\epsilon&1&\chi\cr
-\epsilon\,\chi+f\,\kappa^2\,\psi&\chi&\chi^2-\kappa^2\,f\cr}
\label{matrizg}
\end{equation}
In terms of matrix $h,$ it is possible to write down the field equations (%
\ref{fielde}) in a non-linear $\sigma $-model form 
\begin{equation}
(\rho h_{,z}h^{-1})_{,\overline{z}}+(\rho h_{,\overline{z}}h^{-1})_{,z}=0.
\label{chiral}
\end{equation}
Thus, we can define an abstract Riemannian space using the standard metric
of the group defined by 
\begin{eqnarray}
ds^{2} &=&{\frac{1}{{4}}}tr(dhdh^{-1})  \nonumber \\
&=&{\frac{\rho }{{2\,f^{2}}}}\,[d\,f^{2}+(d\,\epsilon -\psi \,d\chi )^{2}]-{%
\frac{{\rho \,\kappa ^{2}}}{{2\,f}}}\,(d\psi ^{2}+{\frac{1}{{\kappa ^{4}}}}%
\,d\,\chi ^{2})+{\frac{{2\,\rho }}{{3\kappa ^{2}}}}\,d\,\kappa ^{2}
\label{me2}
\end{eqnarray}
This Riemannian space defines a five-dimensional symmetric space (the
covariant derivative of the Riemannian tensor vanishes), with a isometry
group $SL(3,\hbox{{I}\kern-.2em{\bf R}})$.

In what follows we will write explicitly the potentials $\Psi ^{A}$ in terms
of the metric components. In order to do so, we recall that the
five-dimensional space-time metric in terms of the four-dimensional one and
the electromagnetic and scalar fields reads 
\begin{equation}
{ds_{5}}^{2}={\mathaccent94{g}}_{\mu \nu }dx^{\mu }dx^{\nu }+I^{2}(A_{\mu
}dx^{\mu }+dx^{5})(A_{\nu }dx^{\nu }+dx^{5}),  \label{ds5}
\end{equation}
where ${\mathaccent94{g}}_{\mu \nu }$; $\mu ,\nu =1,...,4$ are the
4-dimensional metric components of the five-dimensional space-time, $I$ is
the scalar potential and $A_{\mu }$ is the electromagnetic four potential.
For the axial symmetric stationary case $I,A_{\mu }$ and ${\mathaccent94{g}}%
_{\mu \nu }$ depend only on $\rho $ and $\zeta $. The five-dimensional
metric and its inverse can be written as 
\begin{equation}
{\mathaccent94{g}}_{AB}=\pmatrix{{\mathaccent94{g}}_{\mu\nu}+I^2
A_{\mu}A_{\nu}&I^2 A_\mu\cr I^2A_\mu&I^2\cr}  \label{matrizg1}
\end{equation}
\begin{equation}
{\mathaccent94{g}}^{AB}=\pmatrix{{\mathaccent94{g}}^{\nu\tau}&-A^\nu\cr
A^\nu&A^2+I^{-3}\cr}  \label{matrizg2}
\end{equation}
Due to the symmetries we are working with, its is convenient to write the
four-dimensional metric in the Papapetrou parameterization 
\begin{equation}
{ds_{4}}^{2}=g_{\mu \nu }dx^{\mu }dx^{\nu }={\frac{1}{{f}}}(e^{2k}dzd%
\overline{z}+\rho ^{2}d\varphi ^{2})-f(\omega d\,\varphi +dt)^{2}
\label{me3}
\end{equation}
In terms of this parameterization, and recalling that ${ds_{5}}^{2}={%
\mathaccent94{g}}_{AB}dx^{A}dx^{B}={\frac{1}{{I}}}{ds_{4}}^{2}+I^{2}(A_{\mu
}dx^{\mu }+dx^{5})^{2},$ the metric coefficients ${\mathaccent94{g}}_{AB}$
can be written as 
\begin{equation}
{\mathaccent94{g}}_{AB}=\pmatrix{ 0&{1\over{2If}}e^{2k}&0&0&0\cr
{1\over{2If}}e^{2k}&0&0&0&0\cr 0&0&{\rho^2\over{If}}-{f\omega^2\over{I}}+
{I^2{A_3}^2}&{f\omega\over{I}}+I^2A_3A_4&I^2 A_3\cr
0&0&{f\omega\over{I}}+I^2A_3A_4&-{f\over{I}}+I^2{A_4}^2&I^2A_4\cr
0&0&I^2A_3&I^2A_4&I^2\cr}  \label{matrizg3}
\end{equation}
Using the expressions for the metric given in equation (\ref{me3}), it is
straightforward to calculate the gravitational, electrostatic and scalar
potentials. Recalling that the Killing vectors components $X^{A}$ and $Y^{A}$
satisfy the relations $X^{A}={\delta ^{A}}_{5}$ and $Y^{A}={\delta ^{A}}_{4}$%
, one finds that $Y^{A}Y_{A}={\mathaccent94{g}}_{44}$; and $\ X^{A}Y_{A}={%
\mathaccent94{g}}_{54}.$ Now, substituting these relations into the
definition for $f=-I{\mathaccent94{g}}_{44}+I^{-1}({\mathaccent94{g}}%
_{54})^{2}$ and using the relations (\ref{matrizg3}) we obtain 
\begin{equation}
f=-Ig_{44}
\end{equation}
In similar way one obtains for the electrostatic and the scalar potentials: 
\begin{eqnarray}
\kappa ^{\frac{4}{{3}}} &=&{\mathaccent94{g}}_{55}=I^{2},  \nonumber \\
\psi  &=&-A_{4}.
\end{eqnarray}
For the magnetostatic and rotational potentials the corresponding
expressions can be reduced to 
\begin{eqnarray}
\epsilon _{,\mu } &=&\epsilon _{45\gamma \delta \mu }\mathaccent94{g}^{\tau
\delta }\mathaccent94{g}^{\gamma \theta }\mathaccent94{g}_{4\theta ,\tau }, 
\nonumber \\
\chi _{,\mu } &=&-\epsilon _{54\gamma \delta \mu }\mathaccent94{g}^{\tau
\delta }\mathaccent94{g}^{\gamma \theta }\mathaccent94{g}_{5\theta ,\tau }.
\label{epchi}
\end{eqnarray}
Using now the relations (\ref{matrizg1}), and (\ref{expo}) in (\ref{epchi}),
we find: 
\begin{eqnarray*}
\epsilon _{,\overline{z}} &=&-{\frac{I^{2}}{{\rho }}}[(g_{34}g_{44,\overline{%
z}}-g_{44}g_{34,\overline{z}})]+\psi \chi _{,\overline{z}}, \\
\epsilon _{,z} &=&{\frac{I^{2}}{{\rho }}}[(g_{34}g_{44,z}-g_{44}g_{34,z})]+%
\psi \chi _{z}, \\
\chi _{,\overline{z}} &=&{\frac{I^{4}}{{\rho }}}(g_{34}A_{4,\overline{z}%
}-g_{44}A_{3,\overline{z}}), \\
\chi _{,z} &=&-{\frac{I^{4}}{{\rho }}}(g_{34}A_{4,z}-g_{44}A_{3,z}).
\end{eqnarray*}
The potentials are written in terms of $g_{34}$ and $g_{44}$ components of
the four-dimensional metric tensor as well as of the $A_{3}$ and $A_{4}$
components of the electromagnetic four potential. From (\ref{epchi}) and (%
\ref{matrizg3}) we arrive at the final expressions 
\begin{eqnarray}
A_{3,z} &=&-{\frac{\rho }{{f\kappa }^{2}}}\chi _{,z}+{\frac{g_{34}I}{{f}}}%
\psi _{,z}  \nonumber \\
A_{3,\overline{z}} &=&-{\frac{\rho }{{f\kappa }^{2}}}\chi _{,\overline{z}}+{%
\frac{g_{34}I}{{f}}}\psi _{\overline{z}}  \nonumber \\
\left( {\frac{g_{34}}{{g_{44}}}}\right) _{,z} &=&{\frac{\epsilon _{,z}-\psi
\chi _{,z}}{{f^{2}}}}  \label{compA3yre}
\end{eqnarray}
In the following sections we will use these expressions for obtaining exact
solutions of the field equations.

\section{Calculations and Solutions}

In this section we will apply the previous results for finding exact
solutions of the field equations. Let us consider $f_{0},I_{0},\psi
_{0},..., $ etc. as seed solutions, $i.e.$, as components of the matrix $%
h_{0}$ in (\ref{tinv}). We proceed as follows. First, using the inverse
matrix $h^{-1},$ and the $SL(3,\hbox{{I}\kern-.2em{\bf R}})$ invariance of
the field equations, from (\ref{tinv}) we obtain the $h$ components in terms
of the ${\Psi _{0}^{A}}$ potentials. Finally, using a particular matrix $C$
in (\ref{tinv}) we integrate the new potentials ${\Psi ^{A}}$ in general for
this particular cases.

The inverse matrix of (\ref{matrizg}) reads 
\begin{equation}
h^{-1}=-{\frac{\kappa ^{2/3}}{{f}}}\pmatrix{1&\epsilon-\chi\psi&\psi\cr
\epsilon-\chi\psi&f^2+(\epsilon-\chi\psi)^2-f\chi^2\kappa^{-2}&\kappa^{-2}f%
\chi+\psi(\epsilon-\chi\psi)\cr \psi&\chi
f\kappa^{-2}+\psi(\epsilon-\chi\psi)&-\kappa^{-2}\,f+\psi^2\cr}.
\label{matrizginv}
\end{equation}
Then, we can write the potentials ${\Psi ^{A}}$ in terms of the components
of matrices $h$ and $h^{-1}$ 
\begin{eqnarray}
\kappa ^{\frac{4}{{3}}} &=&{\frac{{h_{11}}^{-1}}{{h_{22}}}};\,\ \ f^{2}={%
\frac{1}{{{h_{11}}^{-1}h_{22}}}};\,\ \chi ={\frac{h_{23}}{{h_{22}}}} 
\nonumber \\
\psi &=&{\frac{{h_{13}}^{-1}}{{{h_{11}}^{-1}}}};\,\ \ \epsilon =-{\frac{%
h_{12}}{{h_{22}}}}  \label{potcomp}
\end{eqnarray}
where we have used the notation ${h_{ij}}^{-1}=(h^{-1})_{ij}$. Using
expressions (\ref{potcomp}) we can straightforwardly calculate the
potentials ${\Psi ^{A}}$ from the $h$ components. We will take each case
separately.

\subsection{Case ${\Psi _{0}}^{A}=(f_{0},0,\protect\psi _{0},0,\protect\kappa%
_{0})$}

In this case we start from a solution with electrostatic, scalar and
gravitational potentials, for this case matrix $h_{0}$ reads 
\begin{equation}
h_{0}=-{\frac{1}{{{\kappa _{0}}^{2/3}f_{0}}}}\pmatrix{{f_0}^2-f_0{%
\kappa_0}^2\psi_0&0&f_0{\kappa_0}^2\psi_0\cr 0&1&0\cr
f_0{\kappa_0}^2\chi_0&0&-\kappa_0^{-2}\,f_0\cr}  \label{matrizg0}
\end{equation}
The inverse matrix is given by 
\begin{equation}
{h_{0}}^{-1}=-{\frac{{\kappa _{0}}^{2/3}}{{f_{0}}}}\pmatrix{1&0&\psi_0\cr
0&f_0^2&0\cr \psi_0&0&{-\kappa_0^{-2}f_0+{\psi_0}^2}\cr}  \label{matrizginv0}
\end{equation}
Now we take the invariance equation $h=ChC^{T}$ taking the constant matrix $%
C $ arbitrary as 
\begin{equation}
C=\pmatrix{a&b&c\cr d&e&j\cr i&h&k\cr}  \label{matrizC}
\end{equation}
and its inverse as 
\begin{equation}
C^{-1}=\pmatrix{q&p&t\cr u&v&w\cr s&y&z\cr}.  \label{matrizCinv}
\end{equation}
Substituting it into (\ref{potcomp}) we arrive at 
\begin{eqnarray}
\kappa ^{\frac{4}{{3}}} &=&{\frac{U}{{V}}}\;\;\;\;\;\;\;\;\;\;\;\;\;\;\;\;\;%
\;\;\;\;\;\;\;\;\;\;\;\;\;\;\;\;\;\;\;\;\;\;\;\;\;\;\;\;\;\;\;\;\;\;\;\;\;\;%
\;\;\;\;\;\;\;\;\;\;\;\;\;\;\;\;\;\;  \nonumber \\
f^{2} &=&{\frac{{f_{0}}^{2}{\kappa _{0}}^{\frac{8}{{3}}}}{{UV}}}%
\;\;\;\;\;\;\;\;\;\;\;\;\;\;\;\;\;\;\;\;\;\;\;\;\;\;\;\;\;\;\;\;\;\;\;\;\;\;%
\;\;\;\;\;\;\;\;\;\;\;\;\;\;\;\;\;\;\;\;\;\;\;\;\;\;\;\;\;\;  \nonumber \\
\chi &=&{\frac{id{f_{0}}^{2}-(id-ij)f_{0}{\kappa _{0}}^{2}{\psi _{0}}%
^{2}+(dk\psi _{0}-kj)f_{0}{\kappa _{0}}^{2}+eh}{{V\kappa _{0}^{-\frac{2}{{3}}%
}}}}  \nonumber \\
\psi &=&-{\frac{{\kappa _{0}}^{2}[-tq-(ts+zq)\psi _{0}-wu{f_{0}}^{2}-sz{\psi
_{0}}^{2}]+szf_{0}}{{U}}}  \nonumber \\
\epsilon &=&-{\frac{f_{0}[daf_{0}+{\kappa _{0}}^{2}{\psi _{0}}^{2}(dc-da)+{%
\kappa _{0}}^{2}(ja\psi _{0}-jc)]+be}{{V\kappa _{0}^{-\frac{2}{{3}}}}}}
\label{poteqC}
\end{eqnarray}
with $U={\kappa _{0}}^{2}(q^{2}+2qs\psi _{0}+u^{2}{f_{0}}^{2}+s^{2}{\psi _{0}%
}^{2})-s^{2}f_{0}$ and $V={\kappa _{0}}^{\frac{2}{{3}}}[d{f_{0}}^{2}-{\kappa
_{0}}^{2}(f_{0}{\psi _{0}}^{2}-2djf_{0}\psi _{0}+j^{2}f_{0})+e^{2}]$. In
order to perform a total integration of the metric components, we take the
matrix $C$ as 
\begin{equation}
C=\pmatrix{1&0&0\cr 0&v&-w\cr 0&-w&v\cr}  \label{matrizC1}
\end{equation}
with inverse 
\begin{equation}
C^{-1}=\pmatrix{1&0&0\cr 0&v&w\cr 0&w&v\cr}  \label{matrizinvC1}
\end{equation}
Then equations (\ref{poteqC}) reduce to the simple expressions 
\begin{eqnarray}
\kappa ^{\frac{4}{{3}}} &=&{\frac{{\kappa _{0}}^{\frac{4}{{3}}}}{{v^{2}-w^{2}%
{\kappa _{0}}^{2}f_{0}}}}  \nonumber \\
f^{2} &=&{\frac{{f_{0}}^{2}}{{v^{2}-w^{2}{\kappa _{0}}^{2}f_{0}}}}  \nonumber
\\
\chi &=&-{\frac{vw(1-f_{0}{\kappa _{0}}^{2})}{{v^{2}-w^{2}{\kappa _{0}}%
^{2}f_{0}}}}  \nonumber \\
\psi &=&v\psi _{0}  \nonumber \\
\epsilon &=&{\frac{wf_{0}\psi _{0}{\kappa _{0}}^{2}}{{v^{2}-w^{2}{\kappa _{0}%
}^{2}f_{0}}}}  \label{ma1s}
\end{eqnarray}
keeping in mind that matrix $C$ fulfills the condition$\,\,\det C=1,$ $i.e.$ 
$v^{2}-w^{2}=1$, from equations (\ref{ma1s}), we obtain the following
relation: 
\begin{equation}
{\frac{\epsilon _{,l}-\psi \chi _{,l}}{{f^{2}}}}={\frac{w{\kappa _{0}}%
^{2}\psi _{0,l}}{{f_{0}}}},  \label{ma1p}
\end{equation}
with $l=z,\overline{z}$. Then, from these last expressions (\ref{ma1p}), and
from the last pair of equations (\ref{compA3yre}), using equation (\ref{ma1s}%
), we have that 
\begin{equation}
\left( {\frac{g_{34}I}{{f}}}\right) _{,z}=-{\frac{w\rho {\kappa _{0}}%
^{2}\psi _{0,z}}{{f_{0}}}};\,\,\,\left( {\frac{g_{34}I}{{f}}}\right) _{,%
\overline{z}}={\frac{w\rho {\kappa _{0}}^{2}\psi _{0,\overline{z}}}{{f_{0}}}}%
.  \label{econzvar}
\end{equation}
We start using the solution (\ref{manq}) and (\ref{manq1}), with $\alpha
^{2}=3$ as seed solution and substituting it into equations (\ref{ma1s}) to
obtain 
\begin{eqnarray}
f^{2} &=&\frac{e^{2\lambda }}{({a_{1}}\tau +a_{2})^{1/2}(v^{2}-w^{2}{{k_{1}}%
^{2}}({a_{1}}\tau +a_{2})e^{2\lambda })}  \nonumber \\
\chi &=&-\frac{vw(1-{{k_{1}}^{2}}({a_{1}}\tau +a_{2})e^{2\lambda })}{%
v^{2}-w^{2}{{k_{1}}^{2}}({a_{1}}\tau +a_{2})e^{2\lambda }}  \nonumber \\
\psi &=&v\frac{{a_{3}}\tau +a_{4}}{{a_{1}}\tau +a_{2}}  \nonumber \\
\epsilon &=&\frac{w{{k_{1}}^{2}}({a_{3}}\tau +a_{4})}{v^{2}-w^{2}{{k_{1}}^{2}%
}({a_{1}}\tau +a_{2})e^{2\lambda }}  \nonumber \\
\kappa ^{4/3} &=&{\kappa _{0}}^{4/3}\frac{({a_{1}}\tau +a_{2})e^{2/3\lambda
})}{v^{2}-w^{2}{{k_{1}}^{2}}({a_{1}}\tau +a_{2})e^{2\lambda }}  \label{red1}
\end{eqnarray}
and substituting this seed solution (\ref{manq}) and (\ref{manq1}) together
with the restriction (\ref{rest1}) into expressions (\ref{econzvar}), we
arrive at: 
\begin{equation}
\left( {\frac{g_{34}I}{{f}}}\right) _{,z}=-wa_{1}k_{1}{\rho }\tau
_{,z};\,\,\,\left( {\frac{g_{34}I}{{f}}}\right) _{,\overline{z}}=wa_{1}k_{1}{%
\rho }\tau _{,\overline{z}}.  \label{econzvar1}
\end{equation}
The integrability of the right hand side of expression ($\ref{econzvar1}$)
is guaranteed because $\tau $ is harmonic and fulfills the Laplace equation (%
\ref{lap}). The explicitly form of the function $g_{34}$ depends on $\tau $.
In \cite{TMRB1}\cite{RBTM} is presented a list of expressions of the rhs of (%
\ref{econzvar}) for different $\tau .$ In terms of $g_{34}$ and the solution
given in (\ref{red1}), we can write the final metric (\ref{ds5}) as

\begin{eqnarray}
{ds_{5}}^{2} &=&{\ \frac{B}{{\kappa _{1}}^{\frac{2}{3}}}}e^{2\kappa -{\ 
\frac{4\lambda }{3}}}dzd\overline{z}+\left( \rho ^{2}e^{-\lambda }{m_{1}}^{%
\frac{1}{2}}-{\frac{\kappa _{1}^{\frac{2}{3}}m_{1}{g_{34}}^{2}}{e^{\frac{%
2\lambda }{3}}}}\right) {\frac{e^{\frac{4\lambda }{{3}}}}{{{\kappa _{1}}^{%
\frac{4}{{3}}}{m_{1}}^{2}}}}d\varphi ^{2}+2g_{34}d\varphi dt  \nonumber \\
&-&{\frac{e^{\frac{2\lambda }{3}}}{{{\kappa _{1}}^{\frac{2}{{3}}}m_{1}}}}%
dt^{2}+\kappa _{1}m_{1}^{\frac{1}{{2}}}e^{\frac{\lambda }{3}}(A_{3}d\varphi
-v{\frac{a_{3}\tau +a_{4}}{{m_{1}}}}dt+dx^{5})^{2}  \label{meso1}
\end{eqnarray}
with 
\begin{equation}
A_{3,z}={\frac{\rho }{{w{\kappa _{1}}^{\frac{2}{{3}}}e^{\frac{4\lambda }{{3}}%
}}}}(\ln {B})_{,z}+{\frac{ve^{\frac{2\lambda }{{3}}}}{{\kappa ^{\frac{5}{{3}}%
}m_{1}}}}g_{34}\tau _{,z}  \label{ds52}
\end{equation}
and $B=v^{2}-w^{2}e^{2\lambda }\kappa _{1}m_{1}$, $m_{1}=a_{1}\tau +a_{2}.$
That is, for a given $\tau ,$ we are able to obtain a rotating exact
solution, with scalar, magnetic and gravitational fields using formulas (\ref
{econzvar}) and (\ref{ds52}).

\bigskip For the second subclass we use the solution (\ref{manq}) with (\ref
{manq2a}) with $\alpha ^{2}=3$ into (\ref{ma1s}) to obtain 
\begin{eqnarray}
f^{2} &=&\frac{e^{2\lambda }}{(a_{1}e^{q_{1}\tau }+a_{2}e^{q_{2}\tau
})(v^{2}-w^{2}{{k_{1}}^{2}}({a_{1}}e^{q_{1}\tau }+a_{2}e^{q_{2}\tau
})e^{2\lambda +\tau _{0}\tau })}  \nonumber \\
\chi &=&-vw\frac{1-{{k_{1}}^{2}}({a_{1}}e^{q_{1}\tau }+a_{2}e^{q_{2}\tau
})e^{2\lambda +\tau _{0}\tau }}{v^{2}-w^{2}{{k_{1}}^{2}}({a_{1}}e^{q_{1}\tau
}+a_{2}e^{q_{2}\tau })e^{2\lambda +\tau _{0}\tau }}  \nonumber \\
\psi &=&v\frac{a_{3}e^{q_{1}\tau }+a_{4}e^{q_{2}\tau }}{a_{1}e^{q_{1}\tau
}+a_{2}e^{q_{2}\tau }}  \nonumber \\
\epsilon &=&\frac{wk_{1}^{2}(a_{3}e^{q_{1}\tau }+a_{4}e^{q_{2}\tau
})e^{2\lambda +\tau _{0}\tau }}{v^{2}-w^{2}{{k_{1}}^{2}}({a_{1}}e^{q_{1}\tau
}+a_{2}e^{q_{2}\tau })e^{2\lambda +\tau _{0}\tau }}  \nonumber \\
\kappa ^{4/3} &=&\kappa _{0}^{4/3}\frac{(a_{1}e^{q_{1}\tau
}+a_{2}e^{q_{2}\tau })e^{2/3(\lambda +\tau _{0}\tau )}}{v^{2}-w^{2}{{k_{1}}%
^{2}}({a_{1}}e^{q_{1}\tau }+a_{2}e^{q_{2}\tau })e^{2\lambda +\tau _{0}\tau }}
\label{solu2}
\end{eqnarray}
In this case, relations (\ref{econzvar}) with the seed solution (\ref{manq})
with (\ref{manq2a}) and the constraints among the constant, equation(\ref
{manq2}), give us 
\begin{eqnarray}
\left( {\frac{g_{34}I}{{f}}}\right) _{,z} &=&-{\frac{w(q_{1}-q_{2})a_{1}%
\kappa _{1}}{\sqrt{a_{1}a_{2}}}}\rho \tau _{,z};  \nonumber \\
\,\,\left( {\frac{g_{34}I}{{f}}}\right) _{,\overline{z}} &=&{\frac{%
w(q_{1}-q_{2})a_{1}\kappa _{1}}{\sqrt{a_{1}a_{2}}}}\rho \tau _{,\overline{z}%
}.
\end{eqnarray}
As in the last case, we can now obtain the line element in terms of the
function $g_{34}$, which can be integrated from these last pair of equations
once a $\tau $ is given. The metric then reads 
\begin{eqnarray}
{ds_{5}}^{2} &=&{\frac{B}{{\kappa _{1}}^{\frac{2}{3}}m_{1}^{\frac{1}{2}}}}%
e^{2\kappa -({\frac{4\lambda }{{3}}}+(q_{1}+q_{2})\tau )}dzd\overline{z}%
+\left( \rho ^{2}\,e^{-\lambda }\,{m_{1}}^{\frac{1}{2}}-{\frac{\kappa _{1}^{%
\frac{2}{{3}}}m_{1}{g_{34}}^{2}}{{e^{{\frac{\lambda }{{3}}}%
+(q_{1}+q_{2})\tau }}}}\right) \times  \nonumber \\
&&{\frac{e^{2{\frac{\lambda }{{3}}}+2(q_{1}+q_{2})\tau }}{{{\kappa _{1}}^{%
\frac{4}{{3}}}{m_{1}}^{2}}}}d\varphi ^{2}+2g_{34}d\varphi dt-{\frac{e^{{%
\frac{\lambda }{{3}}}+(q_{1}+q_{2})\tau }}{{{\kappa _{1}}^{\frac{2}{{3}}%
}m_{1}^{\frac{3}{{2}}}}}}dt^{2}  \nonumber \\
&&+{\kappa _{1}}^{\frac{2}{{3}}}m_{1}e^{\frac{2\lambda }{3}}(A_{3}d\varphi -v%
{\frac{a_{3}e^{q_{1}\tau }+a_{4}e^{q_{2}\tau }}{{m_{1}}}}dt+dx^{5})^{2}.
\label{meso2}
\end{eqnarray}
The function $A_{3}$ again depends on the harmonic function $\tau $ as 
\begin{equation}
A_{3,z}={\frac{\rho }{{\omega \kappa _{1}^{\frac{2}{{3}}}e^{\frac{4\lambda }{%
{3}}}}}}(\ln {B})_{,z}+{\frac{va_{1}(q_{1}-q_{2})e^{{\frac{\lambda }{{3}}}%
+2(q_{1}+q_{2})\tau }}{{\kappa ^{\frac{5}{{3}}}m_{1}}}}g_{34}\tau _{,z}
\end{equation}
where $B=v^{2}-\omega ^{2}e^{2\lambda +(q_{1}+q_{2})\tau }\kappa _{1}m_{1}$
and $m_{1}=a_{1}e^{q_{1}\tau }+a_{2}e^{q_{2}\tau }.$ With solutions (\ref
{meso1}) and (\ref{meso2}) we are now able to obtain rotating exact
solutions which represent rotating monopoles, dipoles, etc. coupled to a
dilaton field.

\subsection{Case ${\Psi _{0}}^{A}=(f_{0},0,0,\protect\chi _{0},\protect\kappa
_{0})$}

We start from a static solution with magnetostatic, scalar and gravitational
potentials. In this case, the matrix $h_{0}$ is

\begin{equation}
h_{0}=-{\frac{1}{{\kappa _{0}^{2/3}f_{0}}}}\pmatrix{f_0^2&0&0\cr
0&1&\chi_0\cr 0&\chi_0&{-\chi_0^2}-{\kappa_0}^{-2}\,f_0\cr}
\label{matrizg04}
\end{equation}
with inverse 
\begin{equation}
{h_{0}}^{-1}=-{\frac{\kappa _{0}^{2/3}}{{f_{0}}}}\pmatrix{1&0&0\cr
0&f_0^2+\chi_0^2\kappa_0^{-2}f_0+f_0^2&f_0 \chi_0 \kappa_0^{-2}\cr 0&f_0
\chi_0 \kappa_0^{-2}&-f_0 \kappa_0^{-2}\cr}.  \label{matrizg04inv}
\end{equation}
Using the invariance equation (\ref{tinv}) and substituting expressions (\ref
{matrizg04}) and (\ref{matrizg04inv}) into the set given by equation (\ref
{potcomp}), with the matrix $C$ given by equation (\ref{matrizC}), we
obtain: 
\begin{eqnarray}
\kappa ^{\frac{4}{{3}}} &=&{\frac{1}{\kappa _{0}^{2/3}}}{\frac{B}{{A}}}%
\;\;\;\;\;\;\;\;\;\;\;\;\;\;\;\;\;\;\;\;\;\;\;\;\;\;\;\;\;\;\;\;\;\;\;\;\;\;%
\;\;\;\;\;\;\;\;\;\;\;\;\;\;\;\;\;\;\;\;\;\;\;\;\;\;\;\;  \nonumber \\
f^{2} &=&{\frac{{f_{0}}^{2}{\kappa _{0}}^{2}}{AB}}\;\;\;\;\;\;\;\;\;\;\;\;\;%
\;\;\;\;\;\;\;\;\;\;\;\;\;\;\;\;\;\;\;\;\;\;\;\;\;\;\;\;\;\;\;\;\;\;\;\;\;\;%
\;\;\;\;\;\;\;\;\;\;\;\;\;\;\;\;\;  \nonumber \\
\chi &=&{\frac{1}{A}}\{di{f_{0}}^{2}+e(h+k\chi _{0})+j(h\chi _{0}+k({\chi
_{0}}^{2}-f_{0}{\kappa _{0}}^{2}))\}  \nonumber \\
\psi &=&{\frac{1}{B}}\{qt{\kappa _{0}}^{2}+uw(-{\chi _{0}}^{2}f_{0}+{\kappa
_{0}}^{2}{f_{0}}^{2})+\chi _{0}f_{0}(sw+uz)-szf_{0}\}  \nonumber \\
\epsilon &=&-{\frac{1}{A}}\{ad{f_{0}}^{2}+be+\chi _{0}(bj+ce)+cj({\chi _{0}}%
^{2}-{\kappa _{0}}^{2}f_{0})\}  \label{poteqC11}
\end{eqnarray}
with $A=d^{2}{f_{0}}^{2}+e(e+2j\chi _{0})+j^{2}({\chi _{0}}^{2}-\kappa
_{0}f_{0})$ and $B=q^{2}{\kappa _{0}}^{2}+u^{2}({f_{0}}^{2}{\kappa _{0}}%
^{2}-f_{0}{\chi _{0}}^{2})+sf_{0}(2u\chi _{0}-s).$ \newline
Next, we take for the matrix $C$ the following particular form: 
\begin{equation}
C=\pmatrix{q&0&-s\cr 0&1&0\cr -s&0&q\cr},
\end{equation}
and its inverse 
\begin{equation}
C^{-1}=\pmatrix{q&0&s\cr 0&1&0\cr s&0&q\cr},
\end{equation}
thus $q^{2}-s^{2}=1$. With this particular form of $C$, the potentials
reduce to the following expressions: 
\begin{eqnarray}
\kappa ^{\frac{4}{{3}}} &=&\kappa _{0}^{4/3}(q^{2}-s^{2}f_{0}{\kappa _{0}}%
^{-2})  \nonumber \\
f &=&\frac{f_{0}}{\sqrt{q^{2}-s^{2}f_{0}{\kappa _{0}}^{-2}}}  \nonumber \\
\chi &=&q\chi _{0}  \nonumber \\
\epsilon &=&s\chi _{0}  \nonumber \\
\psi &=&\frac{sq(1-f_{0}{\kappa _{0}}^{-2})}{q^{2}-s^{2}f_{0}{\kappa _{0}}%
^{-2}}.  \label{cace}
\end{eqnarray}
Using the first differential equation of expressions (\ref{compA3yre}) for $%
A_{3}$ in terms of the seed potentials, (for this case $\epsilon _{0}=\psi
_{0}=0$), we have that 
\begin{eqnarray}
A_{03,z} &=&-\frac{\rho }{f_{0}I_{0}^{3}}\chi _{0,z}  \nonumber \\
A_{03,\bar{z}} &=&\frac{\rho }{f_{0}I_{0}^{3}}\chi _{0,\bar{z}}.
\label{eq:A03}
\end{eqnarray}
On the other hand, from the potentials given in (\ref{cace}), recalling that 
${\kappa _{0}}^{2}={I_{0}}^{3}$, we obtain that in this case: 
\begin{equation}
{\frac{\epsilon _{,z}-\psi \chi _{,z}}{{f^{2}}}}={\frac{s}{f_{0}{I_{0}}^{3}}}%
\,\chi _{0,z}.
\end{equation}
Thus, using the last differential equation from (\ref{compA3yre}) we find
that 
\begin{equation}
-\frac{1}{\rho }\left( \frac{g_{34}}{g_{44}}\right) _{,z}=\frac{s}{%
f_{0}I_{0}^{3}}\chi _{0,z}.
\end{equation}
From equation (\ref{eq:A03}) and this last one, we obtain that 
\begin{equation}
\left( \frac{g_{34}}{g_{44}}\right) _{,z}=sA_{03,z},
\end{equation}
which implies that (up to a constant) 
\begin{equation}
\left( \frac{g_{34}}{g_{44}}\right) =sA_{03},
\end{equation}
that is 
\begin{equation}
g_{34}=g_{44}sA_{03}=\frac{f}{I}sA_{03},
\end{equation}
$i.e.$, for this case we do not need to perform any extra integration for
generating a new rotating solution starting from the seed one. In this way,
we finally obtain the following expression for the target metric 
\begin{eqnarray}
ds_{5}^{2} &=&{\frac{1}{I}}\{T^{\frac{1}{2}}{\frac{e^{k_{0}}}{f_{0}}}dzd%
\overline{z}+\left[ T^{\frac{1}{2}}{\frac{\rho ^{2}}{f_{0}}}-{\frac{s^{2}{%
A_{03}}^{2}f_{0}}{T^{\frac{1}{2}}}}\right] d\varphi ^{2}-{\frac{s{A_{03}}%
f_{0}}{T^{\frac{1}{2}}}}d\varphi dt-{\frac{f_{0}}{T^{\frac{1}{2}}}}dt^{2}\} 
\nonumber \\
&&+I^{2}\left( \frac{qA_{03}}{T}d\varphi -\frac{qs(1-f_{0}{\kappa _{0}}^{-2})%
}{T}dt+dx^{5}\right) ^{2}  \label{mecesar}
\end{eqnarray}
where $T=q^{2}-s^{2}f_{0}{\kappa _{0}}^{-2}$, $%
A_{3}=qA_{03}/(q^{2}-s^{2}f_{0}\kappa _{o}^{-2})$ and $A_{4}=qs(1-s^{2}f_{0}%
\kappa _{o}^{-2})/(q^{2}-s^{2}f_{0}\kappa _{o}^{-2})$. Thus, we have
generated again a new exact solution to the Einstein-Maxwell-dilaton theory,
for $\alpha ^{2}=3$, in which all the fields are non trivially involved.

\subsection{case ${\Psi _{0}}^{A}=(f_{0},\protect\epsilon _{0},0,0,\protect%
\kappa _{0})$}

This case was studied in \cite{TC}, we presented it here somewhat more
detailed in order to have all the cases together. In this case we take as
initial solution one without electrostatic and magnetostatic fields. The
matrix $h_{0}$ is 
\begin{equation}
h_{0}=-{\frac{1}{{\kappa ^{2/3}f_{0}}}}\pmatrix{f_0^2+\epsilon_0^2&-%
\epsilon_0&0\cr -\epsilon_0&1&0\cr 0&0&-{\kappa_0}^{-2}\,f_0\cr}
\label{matrizg01}
\end{equation}
and its inverse is 
\begin{equation}
{h_{0}}^{-1}=-{\frac{\kappa ^{2/3}}{{f_{0}}}}\pmatrix{1&\epsilon_0&0\cr
\epsilon_0&f_0^2+\epsilon_0^2&0\cr 0&0&{-f_0\over{\kappa_0^{-2}}}\cr}.
\label{matrizg01inv}
\end{equation}
In a similar way as in the other cases presented in this work, we take the
equations relating the components of the matrices given in equations (\ref
{matrizg}) and (\ref{matrizginv}), and using (\ref{matrizC}) and (\ref
{matrizCinv}) in the invariance equation (\ref{tinv}), we arrive at 
\begin{eqnarray*}
\kappa ^{\frac{4}{{3}}} &=&{\frac{U}{{{\kappa _{0}}^{\frac{2}{{3}}}V}}} \\
f^{2} &=&{\frac{{f_{0}}^{2}{\kappa _{0}}^{2}}{{UV}}} \\
\chi &=&{\frac{id{f_{0}}^{2}+\epsilon _{0}(id\epsilon _{0}-ie-hd)+he-j\kappa
\kappa _{0}^{2}f_{0}}{{V\kappa _{0}^{\frac{-2}{{V}}}}}} \\
\psi &=&{\frac{{\kappa _{0}}^{2}[tq+wu{f_{0}}^{2}+\epsilon _{0}(uw\epsilon
_{0}+tu+qw)]-szf_{0}}{{U}}} \\
\epsilon &=&-{\frac{f_{0}(daf_{0}-cj\kappa _{0}^{2})+\epsilon
_{0}(ad\epsilon _{0}-bd-ae)+be}{{V}}}
\end{eqnarray*}
where $U=\kappa _{0}(q^{2}+2qu\epsilon _{0}+(uf_{0})^{2}+(u\epsilon
_{0})^{2})-s^{2}f_{0}$, and $V=(df_{0})^{2}+(d\epsilon _{0})^{2}-2de\epsilon
_{0}+e^{2}-(j\kappa _{0})^{2}f_{0}$. In order to integrate the metric, it is
again necessary to consider a simpler matrix $C.$ \ We take 
\begin{equation}
C=\pmatrix{q&0&-s\cr 0&1&0\cr -s&0&q\cr},  \label{matrizC01}
\end{equation}
and its inverse 
\begin{equation}
C^{-1}=\pmatrix{q&0&s\cr 0&1&0\cr s&0&q\cr},  \label{matrizC01inv}
\end{equation}
then, recalling that again $q^{2}-s^{2}=1$, the potentials read 
\begin{eqnarray}
\kappa ^{\frac{4}{{3}}} &=&B{\kappa _{0}}^{\frac{4}{{3}}}  \nonumber \\
f^{2} &=&{f_{0}}^{2}B^{-1}  \nonumber \\
\chi &=&s\epsilon _{0}  \nonumber \\
\psi &=&{\frac{sq[1-\kappa _{0}^{-2}f_{0}]}{{B}}}  \nonumber \\
\epsilon &=&q\epsilon _{0}  \label{poteqC01}
\end{eqnarray}
where $B=q^{2}-s^{2}\kappa _{0}^{-2}f_{0}.$ Integrating equation (\ref
{compA3yre}), and substituting in it expression (\ref{poteqC01}), we obtain: 
\begin{equation}
{\frac{1}{{\rho }}}\left( {\frac{g_{34}I}{{f}}}\right) _{,z}={\frac{q}{{\rho 
}}}\left( {\frac{g_{034}I_{0}}{{f_{0}}}}\right) _{,z},
\end{equation}
which implies that the expression ${\frac{1}{{\rho }}}\left( {\frac{g_{34}I}{%
{f}}}\right) _{,z}$ remains invariant (up to a constant) for the seed
solution and the generated one, $i.e.$

\begin{equation}
g_{34}={\frac{q}{{B}}}g_{034}.
\end{equation}
Similarly we find that 
\begin{equation}
A_{3}=-s{\frac{g_{034}I_{0}}{{B}},}
\end{equation}
$i.e.$, again, for this case we do not need to perform any extra integration
for generating a new rotating solution starting from the seed one.
Substituting the solution into the Papapetrou metric (\ref{ds5}) we arrive
at 
\begin{eqnarray*}
{ds_{5}}^{2} &=&{\frac{1}{{I_{0}f_{0}}}}e^{2k}dzd\overline{z}-B\left[
g_{033}+{\frac{{g_{034}}^{2}I_{0}}{{f_{0}}}}\left( 1-{\frac{q^{2}}{{B}}}%
\right) \right] d\varphi ^{2}+2{\frac{q}{{B}}}g_{034}d\varphi dt \\
&&-{\frac{f_{0}}{{BI_{0}}}}dt^{2}+I^{2}\left( {\frac{-sg_{034}I_{0}}{{\kappa
_{0}^{2}B}}}d\varphi -{\frac{sq[1-\kappa _{0}^{-2}f_{0}]}{{B}}}%
dt+dX^{5}\right) ^{2}
\end{eqnarray*}

Here we must start from a static exact solution coupled to a scalar field.
If there are no extra fields besides the scalar one, the Einstein equations
decouple from the scalar one which satisfy a harmonic equation. The field
equation for the scalar field can be integrated independently from the
Einstein equations. As an exact solution for the scalar field equation we
take the function $\kappa _{0}=\left[ (r-m+\sigma )/({r-m-\sigma })\right]
^{\delta }.$ For the Einstein equations we take as seed metric the Kerr-NUT
space-time, $i.e.$ as seed solution we have 
\begin{equation}
\kappa _{0}=\left( {\frac{r-m+\sigma }{{r-m-\sigma }}}\right) ^{\delta };\,\
\ \epsilon _{0}={\frac{2(\omega L_{+}-lr)}{{\omega }}};\,\ \ f_{0}={\frac{%
\omega -2mr-2lL_{+}}{{\omega }}},
\end{equation}
with 
\begin{equation}
L_{+}=a\cos {\theta }+l;\ \ L_{-}=a\cos {\theta }-l;\,\ \omega =r^{2}+(a\cos 
{\theta }+l)^{2}
\end{equation}
where $r$ and $\theta $ are the Boyer-Lindsquit coordinates, $\rho =\sqrt{%
r^{2}+2mr+a^{2}-l^{2}}\sin {\theta }$ and $\zeta =(r-m)\cos \theta $; $a$, $%
m $ and $l$ respectively are the rotation, mass and NUT parameters, $\sigma $%
, and $\delta $ are integration constants. The resulting target solution is
an exact axial symmetric stationary solution of $5D$ gravity, with
electromagnetic and scalar fields. It reads 
\begin{eqnarray}
{ds_{5}}^{2} &=&{\frac{\omega }{{\omega -2mr-2lL_{+}}}}\left( {\frac{%
r-m+\sigma }{{r-m-\sigma }}}\right) ^{\frac{2\delta }{{3}}%
}(r^{2}-2mr+L_{+}L_{-})e^{2k_{s}}\left( {\frac{dr^{2}}{{\Delta }}}+d\theta
^{2}\right)  \nonumber \\
&&+{\frac{1}{{D}}}\left( {\frac{r-m+\sigma }{{r-m-\sigma }}}\right) ^{\frac{%
4\delta }{{3}}}\{-(\omega
-2mr-2lL_{+})dt^{2}\;\;\;\;\;\;\;\;\;\;\;\;\;\;\;\;\;\;\;\;\;\;\;\;\;\;\;\;%
\;\;\;\;\;\;  \nonumber \\
&&-(4qa(mr+l)\sin ^{2}{\theta }-4l\cos {\theta }\Delta ){\frac{\omega
-2mr-2lL_{+}}{{r^{2}-2mr+L_{-}L_{+}}}}dtd\varphi \;\;\;\;\;\;\;\;\;\;\;\; 
\nonumber \\
&&+[{\frac{\omega }{{\omega -2mr-2lL_{+}}}}\Delta \sin ^{2}{\theta }D\left( {%
\frac{r-m+\sigma }{{r-m-\sigma }}}\right) ^{2\delta
}\;\;\;\;\;\;\;\;\;\;\;\;\;\;\;\;\;\;\;\;\;\;\;\;  \nonumber \\
&&-q^{2}(\omega -2mr-2lL_{+})\left( {\frac{2a\sin ^{2}{\theta }(mr+l)+2l\cos 
{\theta }\Delta }{{r^{2}-2mr+L_{+}L_{-}}}}\right) ^{2}]d\varphi
^{2}\}\;\;\;\;\;\;\;\;\;\;\;\;  \nonumber \\
&&+\left( {\frac{r-m+\sigma }{{r-m-\sigma }}}\right) ^{\frac{2\delta }{{3}}}{%
\frac{D}{{\omega }}}(A_{3}d\varphi
+A_{4}dt+dX^{5})^{2}\;\;\;\;\;\;\;\;\;\;\;\;\;\;\;\;\;\;\;\   \label{meso4}
\end{eqnarray}
where 
\begin{equation}
A_{3}=-\left( {\frac{r-m+\sigma }{{r-m-\sigma }}}\right) ^{2\delta }\left( {%
\frac{2a(mr+l)\sin ^{2}{\theta }+2l\Delta \cos {\theta }}{{%
r^{2}-2mr+L_{+}L_{-}}}}\right) {\frac{\omega -2mr+2lL_{+}}{{D}}}
\end{equation}
\begin{equation}
A_{4}=-qs{\frac{\omega (\frac{r-m+\sigma }{r-m-\sigma })^{2\delta }-(\omega
-2mr+2lL_{+})}{D}}
\end{equation}
\begin{eqnarray*}
D &=&\omega \left( \frac{r-m+\sigma }{r-m-\sigma }\right) ^{2\delta
}q^{2}-s^{2}(\omega -2mr+2lL_{+}); \\
\Delta &=&r^{2}-2mr+a^{2}-l^{2}
\end{eqnarray*}
\begin{equation}
e^{2k_{s}}=\left[ {\frac{(\sqrt{\rho ^{2}+(\zeta -m)^{2}}+\sqrt{\rho
^{2}+(\zeta +m)^{2}})^{2}-4m^{2}}{{4\sqrt{(\rho ^{2}+(\zeta -m)^{2})(\rho
^{2}+(\zeta +m)^{2})}}}}\right] ^{{\frac{8}{{3}}}\delta ^{2}}
\end{equation}
Exact solution (\ref{meso4}) was first obtained in \cite{TC} (see also \cite
{larsen}) and \ contains a great amount of well-known solutions of $5D$
gravity. We will list three of the most important ones. In order to do so,
we start setting $\delta =l=0$ in (\ref{meso4}), obtaining for the 4-dim
space-time (see \cite{TC}) 
\begin{eqnarray}
ds_{4}^{2} &=&{\frac{\sqrt{D\omega }}{{\Delta }}}dr^{2}+\sqrt{D\omega }%
d\theta ^{2}-{\frac{\Delta -a^{2}\sin \theta }{\sqrt{D\omega }}}dt^{2}-{%
\frac{4qamr\sin ^{2}\theta }{\sqrt{D\omega }}}dtd\varphi  \nonumber \\
&&+{\frac{\sin ^{2}\theta }{{\sqrt{D\omega }(\Delta -a^{2}\sin ^{2}\theta )}}%
}[\Delta (D\omega )-4q^{2}a^{2}m^{2}r^{2}\sin ^{2}\theta ]d\varphi ^{2}
\label{mekh}
\end{eqnarray}
where $D=\omega q^{2}-s^{2}(\omega -2mr);$ $\Delta =r^{2}-2mr+a^{2};$ $%
\omega =r^{2}+a^{2}\cos ^{2}\theta .$ For the scalar field the solution
reduces to 
\[
\kappa ^{\frac{4}{{3}}}=q^{2}-s^{2}\left( 1-{\frac{2mr}{{r^{2}+a^{2}\cos
^{2}\theta }}}\right) .\label{ds4f} 
\]

The following known solutions are contained as particular cases of the
generated target solution whose $ds_{4}^{2}$ part is given by equation (\ref
{me3}):

{\bf Frolov-Zelnikov Solution}

This solution is a charged rotating black hole, obtained by Frolov and
Zelnikov in 1987 \cite{FZ}. The $ds_{4}^{2}$ part of the metric is given by 
\begin{eqnarray*}
ds_{4}^{2} &=&-{\frac{1-z}{{B}}}dt^{2}-2a\sin ^{2}{\theta }{\frac{1}{\sqrt{%
1-v^{2}}}}{\frac{z}{{B}}}dtd\varphi \\
&&+\left[ B(r^{2}+a^{2})+a^{2}\sin ^{2}{\theta }{\frac{z}{{B}}}\right] \sin
^{2}{\theta }d\varphi ^{2}+B{\frac{\Sigma }{{\Delta }}}dr^{2}+B\Sigma
d\theta ^{2},
\end{eqnarray*}
where $B=\sqrt{(1-v^{2}+v^{2}z)/({1-v^{2}})}$, $z=2mr/{\Sigma }$, and $%
\Delta =r^{2}+a^{2}-2mr.$ Comparing it with equation (\ref{ds4f}), we find
that this solution corresponds to $\delta =l=0$, $a\neq 0$; $\Sigma =\omega $%
, $D=B^{2}\omega $ and $q=1/\left( {1-v^{2}}\right) $.

{\bf Gibbons-Maeda-Horner-Horowitz Solution}

This solution describes a dilatonic static charged black hole. If we set $%
a=0 $ in (\ref{mekh}) we get the metric 
\begin{equation}
ds_{4}^{2}={\frac{A}{{\Delta }}}dr^{2}-{\frac{\Delta }{{A}}}dt^{2}+r\sqrt{D}%
(d\theta ^{2}+\sin ^{2}\theta d\varphi ^{2})  \label{mehh1}
\end{equation}
where $\omega =r^{2};$ $D=\omega q^{2}-s^{2}(r^{2}-2mr)$; $\ \ \Delta
=r^{2}-2mr$; $\ A=\sqrt{D\omega }.$ Metric (\ref{mehh1}) can be rewritten as 
\begin{equation}
ds^{2}={\frac{1}{f}}dr^{2}-fdt^{2}+r\sqrt{D}d\Omega ^{2}
\end{equation}
where 
\begin{equation}
f={\frac{1-{\frac{2m}{{r}}}}{\sqrt{q^{2}-s^{2}{\frac{1-2m}{{r}}}}}.}
\end{equation}
With the restriction $q^{2}-s^{2}=1,$ the function $f$ transform into 
\begin{equation}
f={\frac{1-{\frac{2m}{{r}}}}{\sqrt{1+{\frac{2m}{{r}}}s^{2}}}}
\end{equation}
If we set $r_{+}-r_{-}=2m;$ \ $r_{-}=2ms^{2};$ $r_{+}=2m(1-s^{2})$ we get $%
f=(1+(r_{-}-r_{+})/r{)/\sqrt{1+r_{-}/r}};$ $r\sqrt{D}=R^{2}=r^{2}\sqrt{%
1+r_{-}/r}$ which corresponds just to the Gibbons-Maeda-Horner-Horowitz
solution \cite{gibb} \cite{HH2}. The charge and the mass parameters can be
written as $Q=ms\sqrt{1-s^{2}};$ $M=m-{\frac{1}{{2}}}ms.$ From the case $s=0$
we find the Schwarzschild solution $(Q=0$ and $M=m).$

Finally, we want to stress the fact that with the procedure of solution
generation presented in this work, the seed solution is also included within
the target solution as a particular case, thus we have:

{\bf Kerr solution}

To recover the Kerr solution setting $\delta =l=0;$ $s=0$, $q=1$ and $a\neq
0 $ in (\ref{mekh}), we get

\begin{eqnarray*}
ds_{4}^{2} &=&-\left( {\frac{\Delta -a^{2}\sin ^{2}\theta }{{\omega }}}%
\right) dt^{2}-{\frac{2a\sin ^{2}{\theta }(r^{2}+a^{2}-\Delta )}{{\omega }}}%
dtd\varphi + \\
&&\sin ^{2}\theta d\varphi ^{2}+{\frac{\omega }{{\Delta }}}dr^{2}+\omega
d\varphi ^{2}
\end{eqnarray*}
with $\Delta =r^{2}-2mr+a^{2}$ and $\omega =r^{2}+a^{2}\cos ^{2}\theta $.

{\bf NUT parameter}

We can also obtain the NUT solution \cite{NUT} taking $a=0$, $\delta =0$ and 
$l\neq 0$ 
\[
ds_{4}^{2}=-{\frac{r^{2}+l^{2}}{{\Delta ^{2}}}}dr^{2}+(r^{2}+l^{2})d\theta
^{2}+({r^{2}+l^{2})\sin }^{2}{\theta }d\varphi ^{2}-{\frac{\Delta }{{%
r^{2}+l^{2}}}}dt^{2} 
\]
$\Delta =r^{2}-2mr-l^{2}$ where $l$ and $m$ respectively are the NUT
parameter and the mass.

\section{Conclusions}

In this work we have given a series of formulae to obtain rotating exact
solutions of the Einstein-Maxwell-Dilaton field equations generated from
seed static ones. The examples we gave for the application of these
formulae, consist on start from a seed solutions in terms of harmonic maps, $%
i.e.$, in terms of two functions which fulfill the Laplace equation. The
static seed solutions represent gravitational fields coupled to a scalar
field and to a magnetostatic (electrostatic) monopoles, dipoles,
quadrupoles, etc. The new solutions generated using our formulae represent
the rotating version of the seed ones. The new solutions contain induced
electric (magnetic) fields generated by the rotation of the body. Some of
the seed solutions model the exterior field of a pulsar containing a scalar
field in the slow rotation limit. The scalar fields could be fundamental or
generated by spontaneous scalarization \cite{dam2}. The new rotating version
of the solution generated using our formulae are in this sense more
realistic and could represent the exterior field of a pulsar with fast
rotation. We suggest that this solutions could be used as theoretical models
for testing the strong gravitational regime of the Einstein theory or the
most important generalizations of general relativity near a pulsar
containing a scalar field. Because of the presence of the electromagnetic
field, our solutions could give also a light to the understanding of such
strong effects like the origin of jets and maybe of the origin of the QPOs,
where approximated and numerical methods could be not completely trustable.

\section{Acknowledgments}

{\hspace{1cm}} This work has been supported by DGAPA-UNAM, project IN105496,
and by CONACYT, Mexico, project 3697E. TM thanks the hospitality from the
relativity group in Jena, Germany, and the DAAD support while this work was
partially done.


\end{document}